\documentclass[conference]{IEEEtran}
\IEEEoverridecommandlockouts
\usepackage{cite}
\usepackage{amsmath,amssymb,amsfonts}
\usepackage{algorithmic}
\usepackage{graphicx}
\usepackage{textcomp}
\usepackage{xcolor}
\usepackage{subfig}
\usepackage{booktabs}
\usepackage{import}
\usepackage{color}
\usepackage[citebordercolor=blue]{hyperref}
\usepackage[capitalise]{cleveref}
\usepackage{multirow}
\usepackage{array}
\usepackage[font=footnotesize]{caption}
\usepackage[inline]{enumitem}
\usepackage{makecell}
\usepackage{siunitx}
\usepackage{caption}
\usepackage{nth}
\usepackage{soul}
\newcommand{\framework}{\mbox{{AstroTimer}}}
\newcommand{\eg}{e.g.,}
\newcommand{\ie}{i.e.,}
\renewcommand\hl[1]{#1} 

\def\BibTeX{{\rm B\kern-.05em{\sc i\kern-.025em b}\kern-.08em
T\kern-.1667em\lower.7ex\hbox{E}\kern-.125emX}}
\author{}
\begin{document}
\bstctlcite{IEEEexample:BSTcontrol}
    \title{AstroTimer: Rethinking Non-Access Stratum Timers in LEO Constellations
    \thanks{We acknowledge the support of the Natural Sciences and Engineering Research Council of Canada (NSERC), [funding reference number RGPIN-2022-03364]}}

    \author{\IEEEauthorblockN{
    Arshiya Rezaie Hezaveh and
    Peng Hu}
\IEEEauthorblockA{
    Advanced Network and Embedded Systems Lab (AEL)\\
    Dept. of Electrical and Computer Engineering, University of Manitoba, Winnipeg, Canada}
    rezaieha@myumanitoba.ca, peng.hu@umanitoba.ca
}

    \maketitle

\begin{abstract}
		Low-Earth Orbit (LEO) constellations expand 5G coverage to remote regions but differ fundamentally from terrestrial networks due to rapidly changing topologies, fluctuating delays, and constrained onboard resources. Existing 3GPP Non-Access Stratum (NAS) timers, inherited from terrestrial and geostationary (GEO) or medium Earth orbit (MEO) systems, fail to accommodate these dynamics, leading to signaling storms and inefficiency. This paper introduces AstroTimer, a lightweight, adaptive framework for sizing NAS timers based on LEO-specific parameters such as link variability, processing delays, and network-function placement. AstroTimer derives a closed-form timer model with low computational cost and optimizes both watchdog and backoff timers for the 5G registration procedure. Simulation results demonstrate that AstroTimer significantly reduces registration time, retry frequency, and user equipment (UE) energy consumption compared to 3GPP defaults, while preventing signaling overloads. The proposed approach provides an operator-ready foundation for reliable, efficient, and scalable non-terrestrial 5G/6G deployments.
    \end{abstract}


    \section{Introduction}\label{sec:introduction}
 
    Non-access stratum (NAS) is a functional layer that includes end-to-end control-plane protocols between the user equipment (UE) and the 5G core network (5GC).
    NAS protocols are essential for handling UE registration, mobility management, and supporting session-related procedures.
    In most NAS protocols, timers are used on both the UE and network function (NF) sides.
    These timers limit waiting periods, detect failures, and trigger reattempts.
    Additionally, by spacing out reattempts, timers can indirectly help regulate the load on the control plane.

    Low-Earth Orbit (LEO) satellite networks (LSNs) are emerging as key enablers of ubiquitous 6G connectivity. Unlike terrestrial networks (TNs) with stable, overprovisioned infrastructure, LSNs face dynamic link conditions, frequent handovers, and variable end-to-end delays due to changing inter-satellite links (ISLs). Moreover, NFs in LSNs can become overloaded with bursty requests, as they are concentrated in a few distant ground gateways, whereas TNs can scale their NFs elastically. Thus, conventional NAS configurations must be redefined to accommodate the unique characteristics of LSNs.
    
    Recent standardization efforts by 3GPP have begun addressing various aspects of non-terrestrial networks (NTN).
    However, NAS timer values have not yet been established for LSNs and remain incomplete for medium Earth orbit (MEO) and geostationary orbit (GEO) systems.
    Unstandardized NAS timers are left to the network operator without guidance on how to estimate them.Relevant research in the literature can be categorized into three areas.
    \begin{enumerate*}
        \item Minimal-change NF deployment: reuse conventional NFs with no or minimal changes~\cite{qosnf,statelessdesignanddeployment}.
        \item NF redesign: merge or split NFs to balance processing load, reduce inter-NF signaling~\cite{onboardsignalingproccessiong,integratedcoresat}.
        \item Signaling reduction: streamline NAS procedures (especially during handovers) to reduce signaling overhead, and prevent or mitigate signaling storms~\cite{handoverpeng, AsynchronousSignaling, 10814619}.
    \end{enumerate*}
    Despite these advances, existing works have hardly addressed timer aspects in LSNs. 

    Despite notable progress in research and standardization, several key questions remain unanswered: Should NAS timers be fixed like TNs? If so, how can fixed values cope with time-varying multi-hop paths and fluctuating control-plane loads without over-provisioning, and do they work in heterogeneous onboard resources and future on-orbit NF deployments? If not, how can timers be dynamically adjusted?
    Our measurements show that using fixed NAS timers causes overhead on NFs and inefficient resource utilization.
    This motivates us to propose an adaptive approach tailored for LSNs.

    
    To address gaps in standardization and research, we introduce \framework{}, which considers the dynamics of LSNs and is ready for both current and future deployments.
    The main contributions of the paper are summarized as follows:
    \begin{itemize}
        \item We derive a tight, closed-form expression for NAS timers with $\mathcal{O}(N)$ complexity. \framework{} captures LEO constellation dynamics and NF placement to produce timer values adaptively.
        \item We instantiate registration timers and evaluate \framework{} across varying UE scales and link-loss regimes using extensive stress simulations on LEO constellations snapshots. Compared to 3GPP defaults, our tuned timers reduce control-plane load through efficient retries and load-aware backoff, shorten failure-detection latency by accounting for link conditions, and decrease UE energy consumption.
        \item \framework{} supports gateway-anchored LSNs and generalizes to future constellations with on-orbit and cross-orbit NF deployments.
    \end{itemize}

    The rest of the paper is structured as follows: Section II discusses the related work. Section III presents the proposed system model and analysis. Section IV discusses the performance evaluation. Section V concludes the paper and outlines the future work.
    \section{Related Work}\label{sec:related-works}
    LEO satellites have fundamental characteristics that distinguish them from terrestrial 5G networks.
    Thus, conventional 5G network design and configuration approaches must be adapted to provide carrier-grade services. In this section, we briefly discusses the 3GPP standardization efforts and relevant works.
    
    In terrestrial 5G networks, significant research has focused on NF scalability~\cite {alawe2018scalability,NFdependency5g},
    signaling load reduction~\cite{waheed2021signalling}, performance analysis during signaling storms~\cite{pavloski2017performance}
    mitigating denial of service signaling threats~\cite{ettiane2021mitigating}.
    Further, authors in~\cite{signaling6g} proposed a new service-based architecture for future 5G/6G networks, where the Access and Mobility Management Function (AMF) no longer
    relays between the UE and NFs. The goal of this approach is to reduce the load on the AMF and increase the network scalability.
    A survey of signaling-related issues can be found in~\cite{signalingSurvey}.

	Research on LSNs has been driven by the  commercial deployments of LEO satellite Internet such as Starlink, OneWeb, and Telesat, and the standardization gap, as early 3GPP NTN work largely reused terrestrial configurations.
    Relevant studies fall into NFs, signaling, and the radio link layer.
    Selected NTN work on core network functions and signaling includes:
    \begin{enumerate*}
        \item \textit{Optimal network function deployment}: Given the resource and power-constrained LEO satellites, core network functions must be carefully deployed on satellites.
        Works in this category aim to reduce overhead and latencies by strategically placing NFs with the least deviation from their standard implementation and signaling procedures~\cite{optimalNFPlacement,qosnf,statelessdesignanddeployment}.

        \item \textit{Core network redesign}: In this category, NFs might merge to reduce inter-NF communication and data redundancy, or an NF might decompose to split processing load, or be redesigned.
        However, the inter-NF communication mechanisms and signaling mainly remain untouched.
        In~\cite{onboardsignalingproccessiong,integratedcoresat}, a framework is proposed to categorize into time-sensitive and time-tolerant, then process time-sensitive signaling via local processing on satellites.
        To mitigate signaling storms and state migration between satellites, a stateless core network is suggested ~\cite{statelessCore}.

        \item \textit{Signaling redesign}: This category of works goes one step further; they modified current signaling procedures or designed ad-hoc signaling from scratch.
        To mitigate signaling storms \mbox{\num{e4} \text{signaling/s per satellite}}~\cite{statelessCore} caused by frequent handovers,
        a conditional handover scheme~\cite{handoverpeng} 
        was suggested.
        Authors in~\cite{AsynchronousSignaling} proposed an asynchronous signaling approach to mitigate the latency and complexity of stateful, synchronous 3GPP signaling. While not explicitly targeting NTNs, it offers a potential solution for such networks.
    \end{enumerate*}

	Recent research has also explored timer and buffer optimization in the 5G radio link layer. In~\cite{layer2timers}, the authors highlight that variable and long round-trip time (RTT) in NTNs necessitate re-evaluating layer-2 timers and buffer sizes. Building on this,~\cite{qosnf} applies a cross-layer approach to adapt timers and buffers according to quality-of-service (QoS) requirements in the service data adaptation protocol.

    Although prior efforts have enhanced NTN signaling efficiency, none have explicitly investigated NAS timer adjustments (e.g., T3510) in LEO constellations as a direct mechanism for signaling load control or faster failure detection. Current NAS timer configurations are predominantly inherited from TNs or scaled based on static RTT estimations, without considering the dynamic nature of LEO satellite networks. Even 3GPP leaves timer settings for LEO deployments undefined in \cite{NASTimers5G}, providing explicit recommendations only for MEO and GEO scenarios.
    
    Additionally, 3GPP delegates the configuration of certain NAS timers (e.g., to network operators), revealing a key research gap and an opportunity for optimization tailored to LEO deployments. Our proposed \framework{} is the first to systematically study and formulate NAS timers specifically for LEO constellations. We analyze and adjust these timers for LSNs and evaluate their impact on failure detection latency and signaling load in an LSN-integrated 5G network.
    \section{\framework{} System Model}\label{sec:system-model}
    \begin{figure}[t]
        \centering
        \includegraphics[width=\columnwidth]{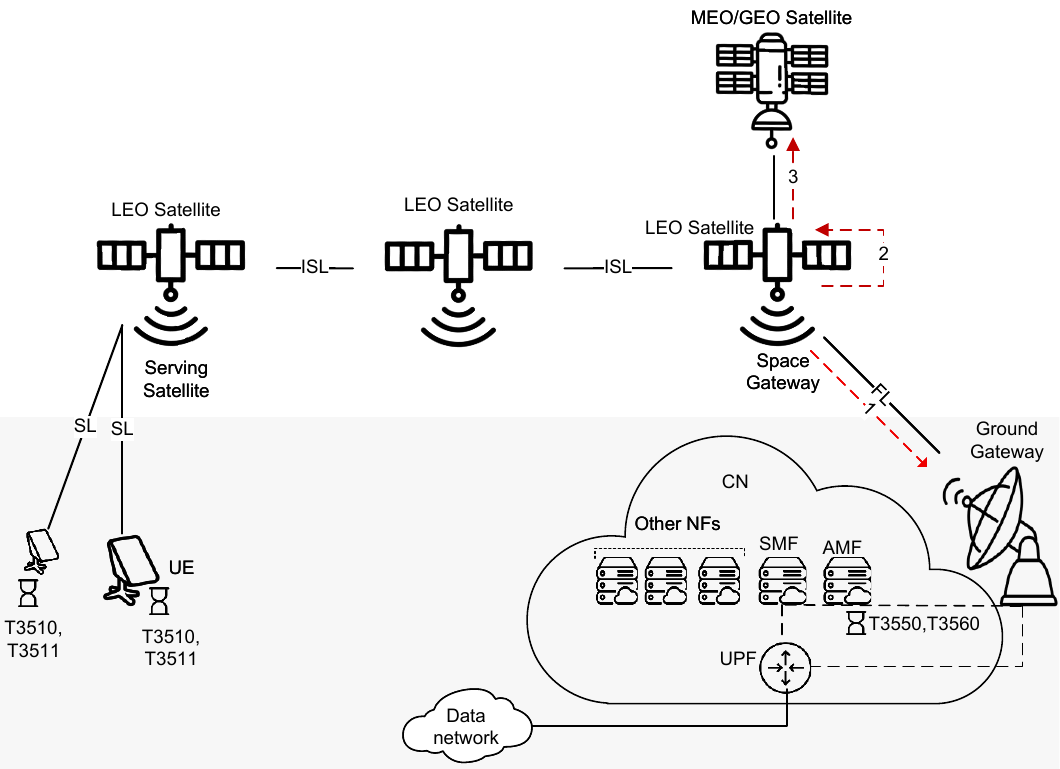}
        \caption{Architecture of an LSN}
        \vspace{-20pt}
        \label{fig:astrotimer_arch}
    \end{figure}        
    \subsection{Overview}\label{subsec:astrotimer-overview}
     This section describes the operation of \framework{}. We first provide a background, then outline the key factors influencing NAS timer adjustment. Next, we categorize the timers and define AstroTimer’s scope, followed by the formulation of its mathematical model.

    \subsection{Background}\label{subsec:backgoround}
    \cref{fig:astrotimer_arch} shows the architecture of an LSN based on 3GPP TR 21.918 in Rel-18.
    Each UE connects to a serving satellite via a service link (SL), which may have ISLs to other LEO satellites or to MEO/GEO satellites in higher orbits, or a feeder link (FL) to the ground gateway.
    For simplicity, cross-orbit ISL is only shown on the space gateway (SG).
    If the serving satellite lacks an FL connection, it forwards traffic to the SG through ISLs.
    The SG then connects to the ground gateway via the FL.
    The ground gateway forwards NAS traffic to the access and mobility function (AMF) and access stratum traffic to the User Plane Function (UPF), which connects to the data network.
    The session management function (SMF) configures the UPF based on QoS requirements.
    NAS timers regulate the progression of NAS procedures (or call flow).
    \cref{fig:astrotimer_arch} also shows the NAS timers used during registration and where each runs, \eg{} T3510 runs on the UE and limits the time to complete registration.

        %
        Registration is the first call flow a UE executes when it powers on.
        Successful registration allows the UE to request additional services.
        The UE begins this process by running T3510 and sending an NAS registration request message to the AMF.
        The AMF then authenticates the UE with the help of other NFs.
        Consequently, the AMF sends an NAS authentication request message to the UE and immediately starts T3550, which limits the time for the UE to respond with an NAS authentication response message.
        The UE can request more services through the NAS registration request message, and the AMF processes these requests after successful authentication.
        Finally, if registration is completed successfully, the AMF runs T3560 and sends an NAS registration accept message to the UE.
        Then, the UE acknowledges the AMF by sending an NAS registration complete message, which the AMF must receive before T3560 expires.

    \subsection{Requirements}\label{subsec:requirements}
        Considering the distinct characteristics of LSNs, we briefly mention the desired properties of an NAS timer configuration framework:
        \begin{enumerate*}
            \item Topology Agnostic: Unlike the terrestrial mobile networks with fixed high-capacity fronthaul links~\cite{functionalSplitTerrestrial}, in LSNs,
            ISLs \hl{in Fig. 1} are made on demand, and the number of intermediate hops may even reach 40~\cite{numisl}.
            Time-varying intermediate hop counts influence communication delay between the UE and NFs, especially the AMF \hl{as shown in Fig. 1}.
            \item NF Deployment Independent: As shown in~\cref{fig:astrotimer_arch}, NFs may be placed (i) on the terrestrial segment behind gateways, simplifying integration but incurring space–ground RTTs and bottlenecks; (ii) on-orbit within the LEO constellation—reducing RTT and detection delay but requiring state mobility and operating under tight resource constraints; or (iii) on higher-orbit (MEO/GEO) satellites, offering stability and wide coverage but inflating RTT. Hybrid deployments (e.g., AMF in orbit, others on ground) are emerging, necessitating timer adaptation to available resources.
            \item Early Failure Detection: ISLs exhibit greater link variability and lower reliability than terrestrial networks due to fast-moving LEO satellites, causing large Doppler shifts and rapid propagation delay changes. While adjusting NAS timers cannot eliminate these issues, it can enable earlier loss detection. For instance, prior work~\cite{layer2timers} refined radio link timers and buffers to account for ISL variability.
            \item Resource Aware: LEO satellites with constrained compute resources may introduce increased per-node latency.
            Prior works \cite{pernodedelay10_2,pernodedelay10}
            report various processing and queueing delay up to \SI{10}{\milli\second}.
            Thus, the proposed framework should not impose computation overhead.
            \item Prevents Signaling Storm: LSNs are prone to signaling storms. The short visibility window of LEO satellites forces UEs to connect to the nexta handover), resulting in(\ie{} handover), generating a large volume of signaling. If timers are not set accurately, consecutive NAS messages may be sent, further increasing the signaling load on the satellite.
        \end{enumerate*}
        
        Designing \framework{} to set NAS timers appropriately involves balancing conflicting objectives. Longer timers help prevent signaling storms, while shorter ones enable faster loss detection. To reconcile this trade-off, we introduce adjustment coefficients ($\alpha$ and $\beta$) that balance early reattempts with adequate waiting periods to mitigate bursty load. The detailed design of \framework{} is discussed in~\cref{subsec:mathematical-modeling}.

    \subsection{Categorizing Timers}\label{subsec:design}
        NAS timers can be divided into three categories:
        \emph{Watchdog} timers determine the maximum time to receive the expected response(s), \eg{} T3510 governs the time to register the UE.
        \emph{Backoff} timers prevent the initiation of a procedure until expiration, \eg{} T3511 prevents a UE from making more registration attempts until expiration. A failure may cause a backoff timer to run.
        \emph{Periodic} timers indicate performing a job on expiration, \eg{} T3512 indicates the time to perform periodic reregistration. We design \framework{} to adjust \textit{watchdog} and \textit{backoff} timers.

    \subsection{Mathematical Modeling}\label{subsec:mathematical-modeling}
\begin{table}
	\caption{\framework{} Variables}\label{tab:astro_variables}
	\centering
	\renewcommand{\arraystretch}{1.15}
	\resizebox{\columnwidth}{!}{
		\begin{tabular}{l l}
            \toprule
			\textbf{Variable} & \multicolumn{1}{c}{\textbf{Description}} \\ 
            \midrule
			$D^{i}_{\mathrm{agg}},\text{ }D_{\mathrm{ss}}^{i},\text{ },D_{\mathrm{brs}}^{i}$    & Aggregated, steady-state(ss), and burst delays in \emph{i} \\
			$D_{\mathrm{prp}}^{\left< i,j \right>}$                                             & Propagation delay on the link between \emph{i} and \emph{j} \\
			$\mu^{i}$                                                                             & Processing rate \\
			$\lambda^i, \ \lambda_{\mathrm{ss}}^{i}, \ \lambda_{\mathrm{brs}}^{i}$              & Total/steady state/burst packet(or request) arrival rate\\
			$t_{\mathrm{brs}}^{i}$                                                              & Burst duration \\
			$\alpha,\beta \in \left( 0,1 \right]$                                                          & Adjustment coefficients \\
			$R$                                                                                 & The number of handshakes in a NAS procedure \\
			$N$                                                                                 & \makecell[l]{ Hop-count,\ie{} number of intermediate nodes between \\ the timer origin and the responder } \\
            $T$                                                                                 & \framework{}-computed NAS timer value  \\
            \bottomrule
		\end{tabular}
	}
	\caption*{
    \scriptsize {Note: Superscripts denote entities, and subscripts denote event types. For example, $\lambda^{i}_{\mathrm{brs}}$ shows burst arrival rate at $i$.}
    }
	\vspace{-15pt}
\end{table}
        Here we present the proposed closed-form expression for calculating NAS timers as shown in (\ref{eq:astro_timer}).
        \cref{tab:astro_variables} shows variables used.
        
        \hl{The input is the path from the origin (\ie{} the entity that runs the timer) to the responder (\ie{} the entity to which the origin communicates).}
        The first term in (\ref{eq:astro_timer}) accounts for the end-to-end delay, except at the endpoint nodes.
        For each hop $i$, the total delay is $D_{\mathrm{agg}}^{i}$, capturing steady-state ($D_{\mathrm{ss}}^{i}$) and burst-induced backlog ($D_{\mathrm{brs}}^{i}$),
        and $D_{\mathrm{prp}}^{\left< i,i+1 \right>}$ accounts for propagation between consecutive nodes.
        \begin{subequations}
            \begin{flalign}
                & T\!= R (D_{\mathrm{prp}}^{\left< 0,1 \right>} + \sum_{i=1}^{N-1} ( D^{i}_{\mathrm{agg}} + D_{\mathrm{prp}}^{\left< i,i+1 \right>})+ D_{\mathrm{prp}}^{\left< N-1,N \right>})&&&
                \notag\\
                &  \phantom{=}+(\lfloor \frac{R}{2} \rfloor+1).(  \alpha D^{\mathrm{0}}_{\mathrm{agg}} + \beta D^{\mathrm{N}}_{\mathrm{agg}}) \label{eq:astro_timer}\\
                & D^{i}_{\mathrm{agg}} = D_{\mathrm{ss}}^{i} + D_{\mathrm{brs}}^{i}\label{eq:delay_components}&&&\\
                & D_{\mathrm{ss}}^{i}\hphantom{D} {=} \frac{1}{\mu-\lambda_{\mathrm{ss}}} \label{eq:ss_delay}&&&\\
                & D_{\mathrm{brs}}^{i}\!\hphantom{D} {\!=} \frac{(\lambda_{\mathrm{brs}}-\mu)t_{\mathrm{brs}}}{\mu}\label{eq:ins_delay}&&&\\
                & \lambda_\mathrm{brs}^{i} \hphantom{=} {=}
                \begin{cases}
                    0 & \lambda < \mu \\
                    \lambda - \lambda_{\mathrm{ss}} & \lambda \geq \mu
                    \label{eq:burst_detection}
                \end{cases}
            \end{flalign}
        \end{subequations}
\addtolength{\topmargin}{0.03in}
        The second term in~\eqref{eq:astro_timer} aggregates endpoint processing at the origin ($i=0$) and the responder ($i=N$) across  handshakes:
        $D_{\mathrm{agg}}^{0}$ and $D_{\mathrm{agg}}^{N}$ denote their per-round delays.
        Endpoint weights $\alpha$ and $\beta$ modulate their contributions across rounds to tune conservativeness.
        Larger weights prevent early timeouts and excessive attempts; however, they may delay the detection of loss. 
        To balance early loss detection against avoiding premature retries, $\alpha$ and $\beta$ must be tuned:
        as shown in~(\ref{eq:delay_components}), $D_{\mathrm{agg}}^{i}$ consists of two components:
        the steady-state delay ($D_{\mathrm{ss}}^{i}$) and the transient delay ($D_{\mathrm{brs}}^{i}$) caused by bursts.
        Bursts must be treated carefully because they cause a significant increase in the aggregated delay.
        As shown in (\ref{eq:burst_detection}), if $\lambda>\mu$ means the arrival rate is greater than the processing rate, which causes a queue during $t_{\mathrm{brs}}$.
        Otherwise, $D_{\mathrm{brs}}^{i}=0$.

        For a given topology and fixed $R$, computing $T$ takes linear time in the number of intermediate hops, \ie{} the time complexity is $\mathcal{O}(N)$.
        Computing $D_{\mathrm{agg}}^{i}$ and $D_{\mathrm{prp}}^{\left< i,i+1 \right>}$ takes constant time. 
        Thus, each hop contributes $\mathcal{O}(1)$; over \emph{N} hops, the total is $\mathcal{O}(N)$.
        Endpoint-delay computations are independent of \emph{N} and take $\mathcal{O}(1)$ time, so the overall complexity remains \(\mathcal{O}(N)\).

    \subsection{Real-World Example}\label{subsec:realworld-example}


        Now, we explain how \framework{} calculates timers.
        According to our definition, T3510, T3550, and T3560 are \emph{watchdog} timers, and T3511 is a \emph{backoff} timer.
        The UE starts the registration procedure and exchanges NAS messages with the AMF over five rounds, \ie{} $R=5$.
        Therefore, the UE is the origin and the AMF is the responder.
        T3550 and T3560 both run in the AMF (origin) to limit the time to receive a response from the UE (responder).
        

        If T3510 expires or an NAS Registration Reject is received, the UE (origin) initiates the backoff timer T3511, which is adjusted based on the AMF's estimated load (responder). During the backoff period, no messages are exchanged, meaning $R\!=\!0$.
\section{Evaluation}\label{sec:evaluation}

	We implement a Python-based NAS emulator that supports 5G NAS procedures.
	For evaluation, we focus on the registration call flow of the most critical procedures since a UE must register before any service.
	We then compare \mbox{\framework{}'s} adaptive timers with 3GPP reference timers across diverse network conditions and topologies.

	\subsection{Evaluation Environment}\label{subsec:simulation-environment}
\begin{table}
    \caption{Evaluation Parameters}
    \label{tab:evaluation_parameters}
    \centering
    \renewcommand{\arraystretch}{1.2}
    \begin{tabular}{ll|ll}
        \toprule
        \textbf{Property}   & \textbf{Value}    & \textbf{Property}     & \textbf{Value} \\
        \midrule
        $\alpha,\beta$      & 0.5               & Loss probability      & 0--50{\%}~\cite{SaT5Glossprob} \\
        T3510               & 27 s~\cite{NASTimers5G} & T3511                       & 18~s~\cite{NASTimers5G} \\
        T3550         &  11~s~\cite{NASTimers5G} & T3560 &~11 s~\cite{NASTimers5G}\\
        Power (idle)   & 20 W\cite{starlinkpowercons}   & Power (active) & 50--75W~\cite{starlinkpowercons} \\
        $D_{\mathrm{prp}}^{\mathrm{SL}}$, $D_{\mathrm{prp}}^{\mathrm{FL}}$ & \SI{2}{\milli\second} & $D_{\mathrm{agg}}^{\mathrm{UE}}$  & $\approx\!1~$\!ms~\cite{handoverpeng} \\
        $D_{{\mathrm{prp}}}^{\mathrm{ISL}}$ & 16~ms\cite{chaudhry2021laser} &        Avg. hop count           &  16~\cite{hopcount}  \\
        $\mu^{\mathrm{NF}}$ & 300{~req/s} & $\mu^{\mathrm{sat}}$                   & $1.6\times10^7$ {pkt/s} \\
        \bottomrule
    \end{tabular}
    \vspace{-15pt}
\end{table}
		We performed evaluations under stressed conditions.
		A steady-state load at \SI{80}{\percent} is maintained, with superimposed bursty arrivals, each lasting $t_{\mathrm{brs}}=\SI{1}{\milli\second}$.
		We increase the number of UEs from 4K to 5K and the packet-loss probability from \SI{0}{\percent} to \SI{50}{\percent}.
		Results report the average of 10 independent runs per configuration.
		Key parameters can be seen in \cref{tab:evaluation_parameters}.

		Since 3GPP has not specified NAS timers for LSNs, we use MEO/GEO reference values in 3GPP TS 24.501~\cite{NASTimers5G}.
		T3511 remains unspecified even for MEO/GEO~\cite{NASTimers5G}.
		Noticing an approximate $1.8\times$ scaling from terrestrial to MEO/GEO (\eg{} T3510: $27/15\!\approx\!1.8$),
		we set T3511 to $\SI{18}{\second}$ by applying the same factor to the terrestrial default.

	\subsection{Timer Sizing Examples}\label{subsec:timer-examples}
		We size timers for a 4K-UE and zero-loss scenario, where the UE is attached to the last satellite on the path to AMF.
		At each intermediate satellite,~(\ref{eq:ss_delay}) gives a steady-state delay of $D_{\mathrm{ss}}\approx\SI{0.3}{\micro\second}$, which is negligible.
		Because $\lambda_{\mathrm{brs}}=0$ at these nodes, the burst term vanishes ($D_{\mathrm{brs}}=0$).
		At the AMF, the steady-state delay is $D_{\mathrm{ss}}^{\mathrm{AMF}}\approx\SI{200}{ms}$.
		Under a burst arrival rate of $\num{4e3}$ requests/\si{ms} (i.e., $\lambda_{\mathrm{brs}}\gg \mu $), a backlog forms at the AMF\@.
		Applying~(\ref{eq:ins_delay}) yields a burst-induced delay of $D_{\mathrm{brs}}^{\mathrm{AMF}}\approx\SI{12.3}{s}$.

		The path and processing parameters are identical across T3510, T3511, T3550, and T3560.
		The only varying factor is the handshake count $R$.
		We use $R{=}5$ for T3510 and $R{=}2$ for T3550/T3560, while T3511 is a backoff timer with $R{=}0$.
		Substituting the numerical values into~(\ref{eq:astro_timer}) yields $\mathrm{T3510}\approx\SI{24}{s}$, $\mathrm{T3511}\approx\SI{8}{s}$ and $\mathrm{T3550}=\mathrm{T3560}\approx\SI{16}{s}$.

	\subsection{Performance Metrics}\label{subsec:performance-metrics}
		We benchmark \framework{} against 3GPP timers and discuss key performance results.

		\subsubsection{Registration Length CDF}
\begin{figure}[t]
    \centering

    \includegraphics[width=0.75\columnwidth]{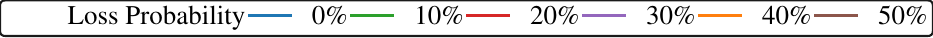} \hfill \\
    \vspace{-3mm}
    \subfloat[\framework{}, 3K UEs]{%
        \includegraphics[width=0.33\columnwidth]{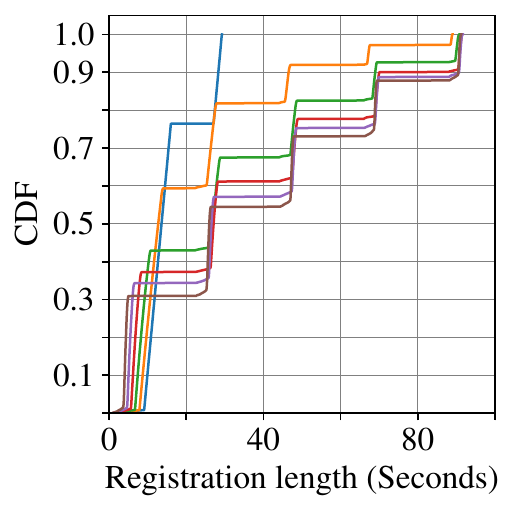}%

        \label{fig:reg_astro_3k}
    }
    \hspace{-3mm}
    \subfloat[\framework{}, 4K UEs]{%
        \includegraphics[width=0.33\columnwidth]{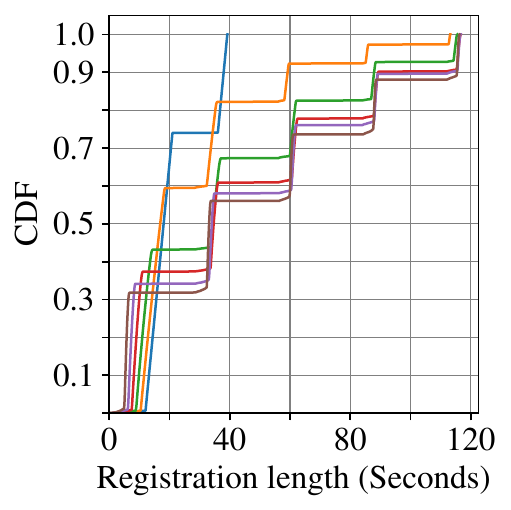}%
        \label{fig:reg_astro_4k}
    }
    \hspace{-4mm}
        \subfloat[\framework{}, 5K UEs]{%
        \includegraphics[width=0.33\columnwidth]{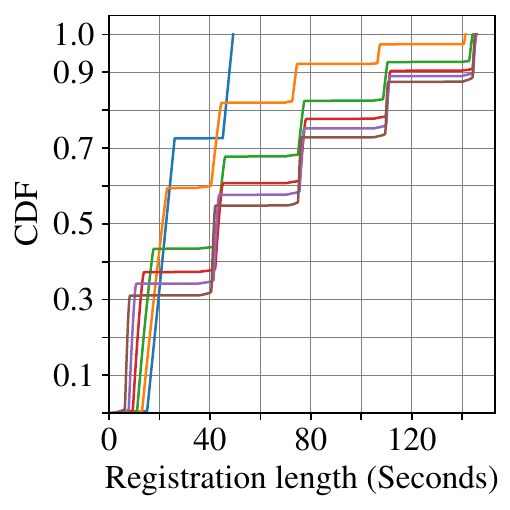}%
        \label{fig:reg_astro_5k}
    } \\[1ex]
    \vspace{-3mm}
        \subfloat[3GPP, 3K UEs]{%
        \includegraphics[width=0.33\columnwidth]{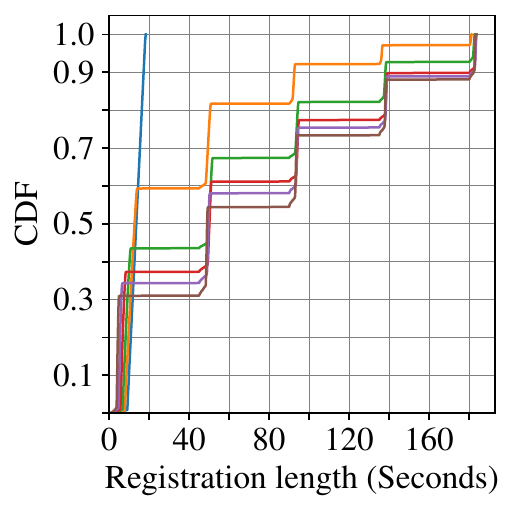}%
        \label{fig:reg_3gpp_3k}
    }
    \hspace{-4mm}
    \subfloat[3GPP, 4K UEs]{%
        \includegraphics[width=0.33\columnwidth]{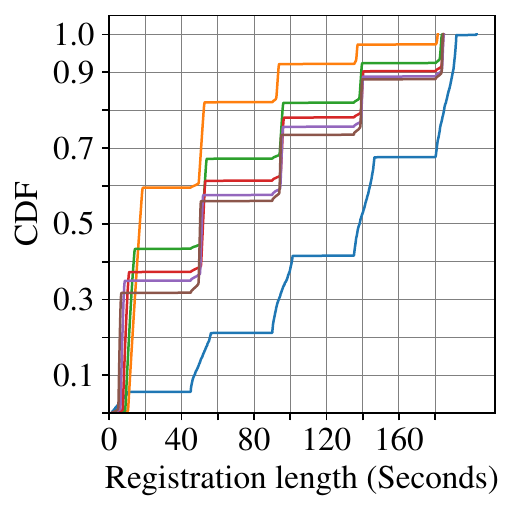}%
        \label{fig:reg_3gpp_4k}
    }
    \hspace{-4mm}
        \subfloat[3GPP, 5K UEs]{%
        \includegraphics[width=0.33\columnwidth]{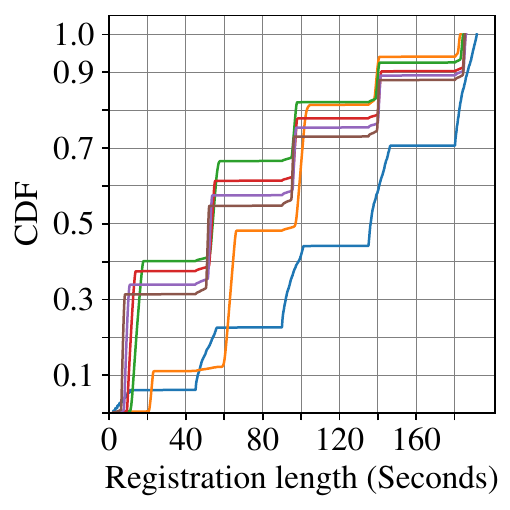}%
        \label{fig:reg_3gpp_5k}
    } \\[1ex]
    \caption{CDF of the registration time. (Left-shifted curves are preferred.)}
    \label{fig:cdf_reg_plots}
	\vspace*{-6mm}
\end{figure}

			\cref{fig:cdf_reg_plots} shows the CDF of registration time for UEs that eventually succeed.
			Each curve aggregates successful registrations regardless of the number of retries.
			Per~\cite{NASTimers5G}, each UE is limited to five attempts.

			With 3K UEs and zero loss, both methods have the same registration time, as shown in \cref{fig:reg_astro_3k,fig:reg_3gpp_3k}.
			The AMF remains responsive, and queues stay short, so no timers expire and no losses occur.
			As the loss probability increases, \framework{} still achieves a lower registration time.
			Under low load, \framework{} sets a smaller value for T3511, i.e., \SI{3}{s}.
			UEs therefore retry earlier, reducing time-to-register.

			With 4K UE (\cref{fig:reg_astro_4k,fig:reg_3gpp_4k}) and 5K (\cref{fig:reg_astro_5k,fig:reg_3gpp_5k}) UE cases, the AMF queue grows long and functions near saturation.
			Nevertheless, \framework{} reduces registration time by increasing timers based on the estimated AMF load.
			In other words, \framework{} adjusts T3511 to prevent backlog bursts on the AMF and to reduce idle time after loss. At zero loss, 3GPP may appear faster in the CDF conditioned on success, yet \SI{97}{\percent} of UEs never register.

		\subsubsection{Expired timers}
\begin{figure}[t]
	\captionsetup[subfloat]{labelfont=tiny,textfont=tiny}
	\centering
	\subfloat[\framework{}, 4K UEs]{%
		\includegraphics[width=0.23\linewidth]{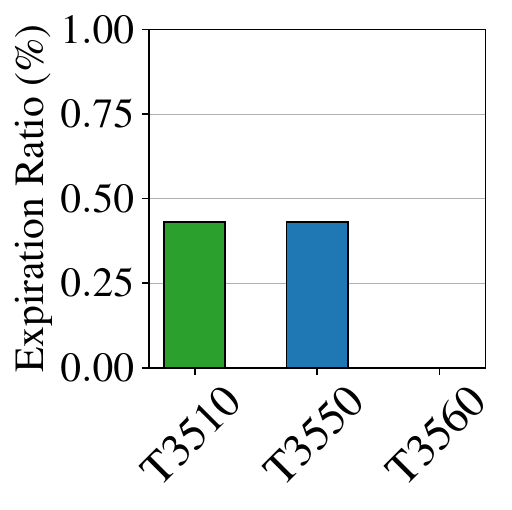}%
		\label{fig:expired_astro_4k}
	}
	\hfill
	\subfloat[\framework{}, 5K UEs]{%
		\includegraphics[width=0.23\linewidth]{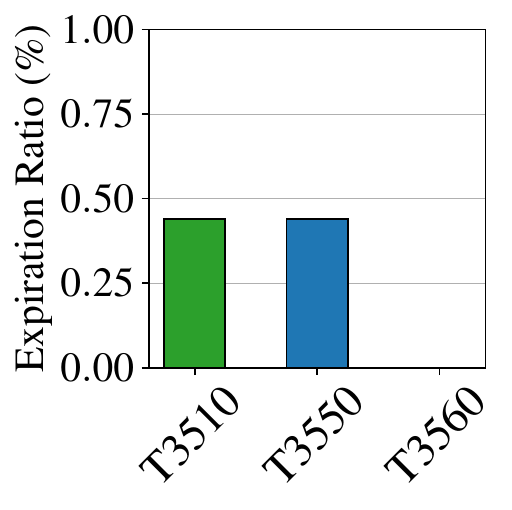}%
		\label{fig:expired_astro_5k}
	}
	\hfill
	\subfloat[3GPP, 4K UEs]{%
		\includegraphics[width=0.23\linewidth]{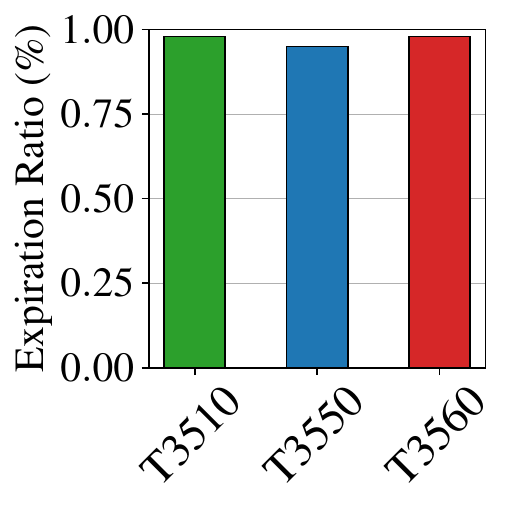}%
		\label{fig:expired_3gpp_4k}
	}
	\hfill
	\subfloat[3GPP, 5K UEs]{%
		\includegraphics[width=0.23\linewidth]{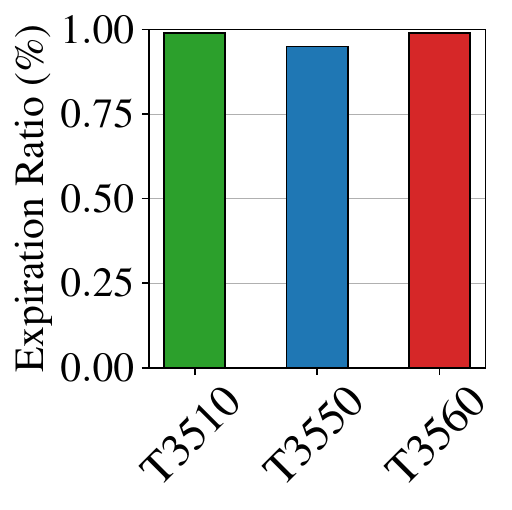}%
		\label{fig:expired_3gpp_5k}
	}
	\caption{Ratio of expired timers to all timers}
	\label{fig:expired_timers}
	\vspace*{-6mm}
\end{figure}

			\cref{fig:expired_timers} shows the ratio of expired to all timers.
			In other words, timer expiration indicates that the timer value is insufficient, which causes further attempts.

			We intentionally used small $\alpha$ and $\beta$, so approximately \SI{20}{\percent} of timers expired, whereas overprovisioned 3GPP timers did not.
			This tradeoff comes at the cost of more tries when the loss probability is zero, but reduces registration time when the loss probability is nonzero.
			With 4K and 5K UEs and \framework{} timers, all UEs could eventually register.
			However, nearly \SI{40}{\percent} of T3510 and T3550 timers, and none of the T3560 timers, expired.
			With 3GPP timers, about 99\% of timers expired.
			This led to failed registrations for more than 96\% UEs.

		\subsubsection{Registration attempts}
\begin{figure}[t]
    \centering

    \includegraphics[width=0.75\columnwidth]{Figures/legend} \hfill \\
    \vspace{-3mm}
    \subfloat[\framework{}, 3K UEs]{%
        \includegraphics[width=0.33\columnwidth]{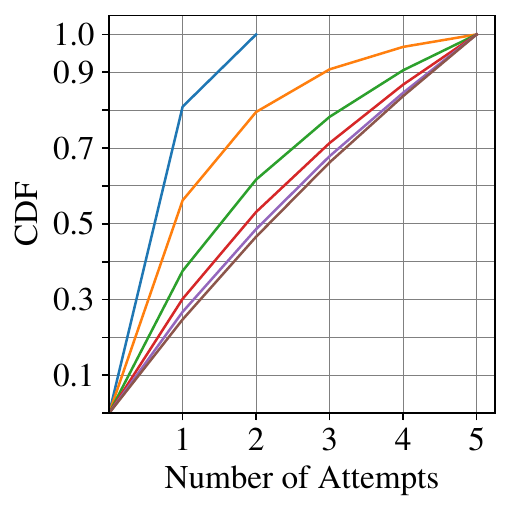}%

        \label{fig:reg_attempts_astro_3k}
    }
    \hspace{-3mm}
    \subfloat[\framework{}, 4K UEs]{%
        \includegraphics[width=0.33\columnwidth]{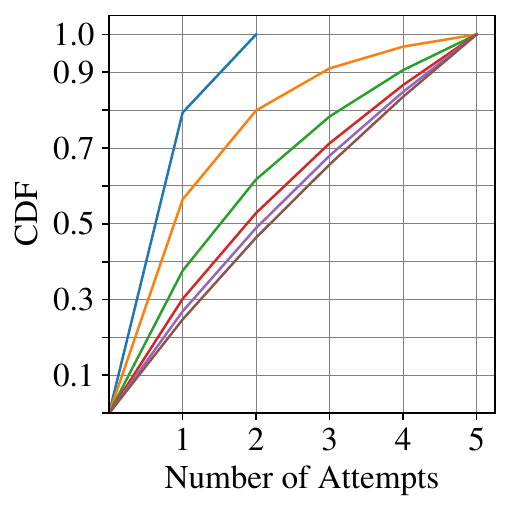}%
        \label{fig:reg_attempts_astro_4k}
    }
    \hspace{-3mm}
        \subfloat[\framework{}, 5K UEs]{%
        \includegraphics[width=0.33\columnwidth]{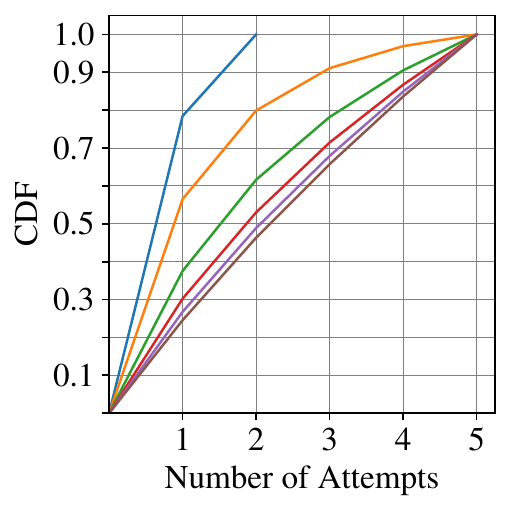}%
        \label{fig:reg_attempts_astro_5k}
    } \\[1ex]
    \vspace{-2mm}
        \subfloat[3GPP, 3K UEs]{%
        \includegraphics[width=0.33\columnwidth]{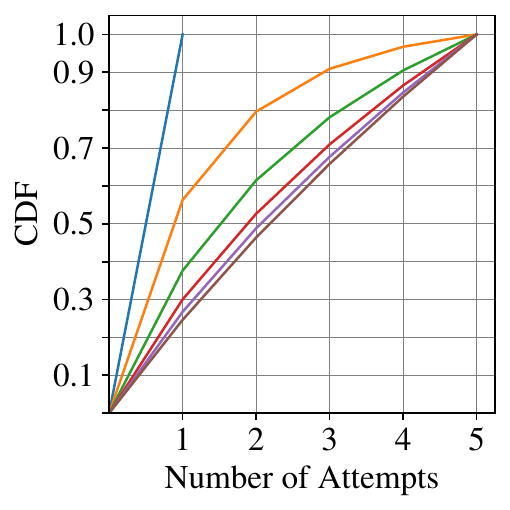}%
        \label{fig:reg_attempts_3gpp_3k}
    }
    \hspace{-3mm}
    \subfloat[3GPP, 4K UEs]{%
        \includegraphics[width=0.33\columnwidth]{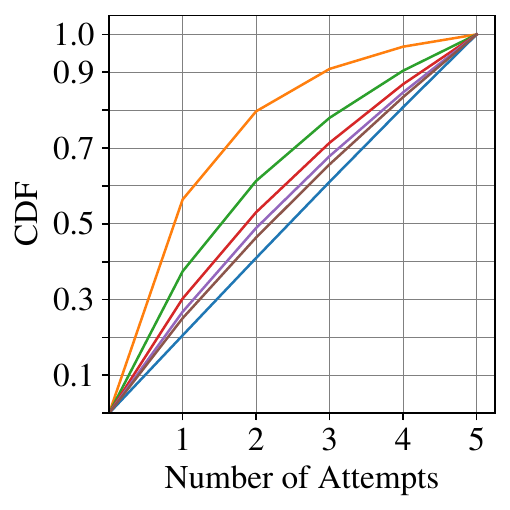}%
        \label{fig:reg_attempts_3gpp_4k}
    }
    \hspace{-3mm}
        \subfloat[3GPP, 5K UEs]{%
        \includegraphics[width=0.33\columnwidth]{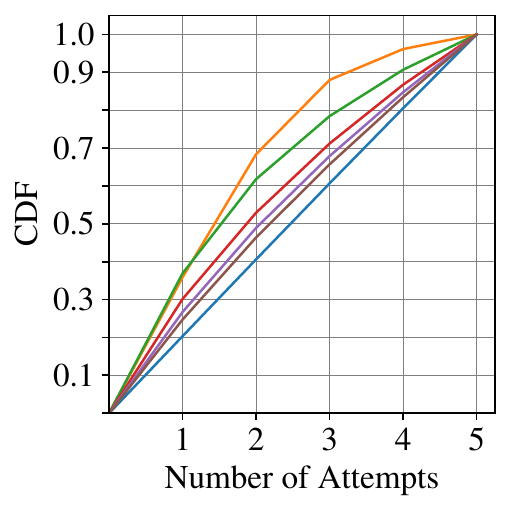}%
        \label{fig:reg_attempts_3gpp_5k}
    } \\[1ex]
    \caption{CDF of the number of registration attempts with zero loss probability.}
    \label{fig:cdf_reg_attempts_plots}
	\vspace*{-3mm}
\end{figure}

			\cref{fig:cdf_reg_attempts_plots} shows how many times UEs attempted to register.
			It is preferred to register devices with the fewest attempts and the least time.
			It can be seen that \framework{} always has fewer registration attempts except with 3K UEs.
			We intentionally decreased $\alpha$ and $ \beta$, so T3550 and T3510 expired in some cases, leading to more attempts.
			It is worth noting that when the loss probability is zero, with \framework{} timer, all UEs successfully register.
			In 3GPP, with 4K and 5K UEs, almost \SI{96}{\percent} UEs fail to register even after five attempts.
		\subsubsection{CDF energy consumption}
\begin{figure}[t]
    \centering

    \includegraphics[width=0.75\columnwidth]{Figures/legend} \hfill \\
    \vspace{-3mm}
    \subfloat[\framework{}, 3K UEs]{%
        \includegraphics[width=0.33\columnwidth]{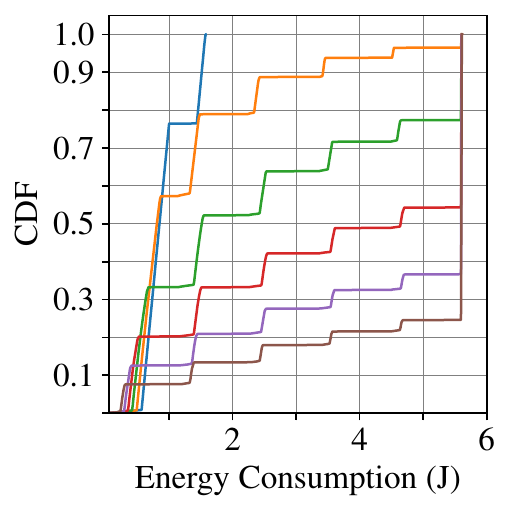}%

        \label{fig:energy_all_astro_3k}
    }
    \hspace{-3mm}
    \subfloat[\framework{}, 4K UEs]{%
        \includegraphics[width=0.33\columnwidth]{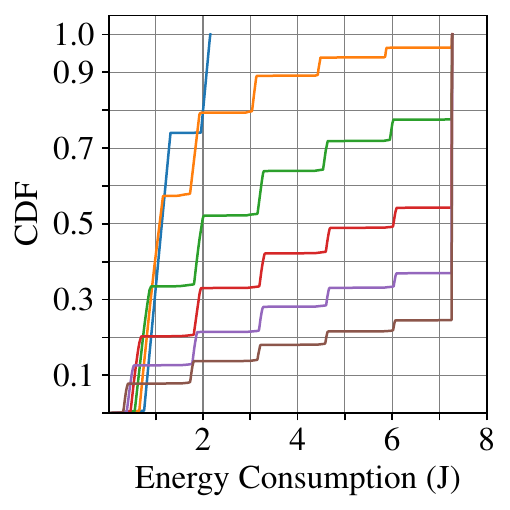}%
        \label{fig:energy_all_astro_4k}
    }
    \hspace{-3mm}
        \subfloat[\framework{}, 5K UEs]{%
        \includegraphics[width=0.33\columnwidth]{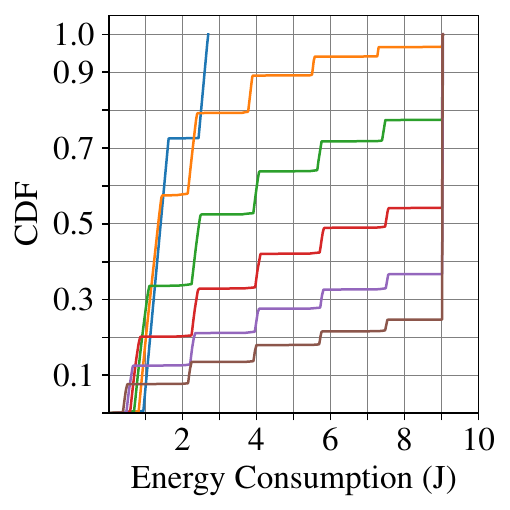}%
        \label{fig:energy_all_astro_5k}
    } \\[1ex]
    \vspace{-2mm}
        \subfloat[3GPP, 3K UEs]{%
        \includegraphics[width=0.33\columnwidth]{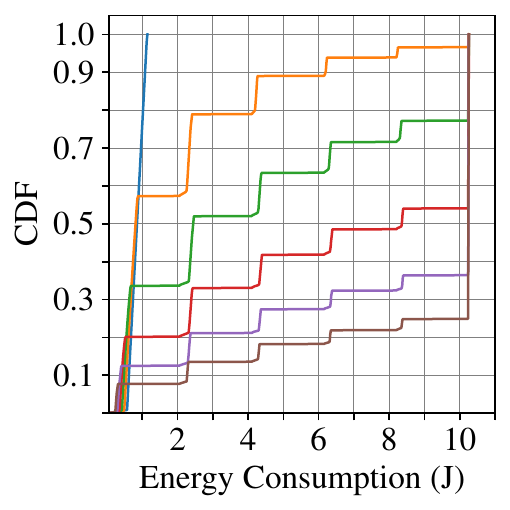}%
        \label{fig:energy_all_3gpp_3k}
    }
    \hspace{-3mm}
    \subfloat[3GPP, 4K UEs]{%
        \includegraphics[width=0.33\columnwidth]{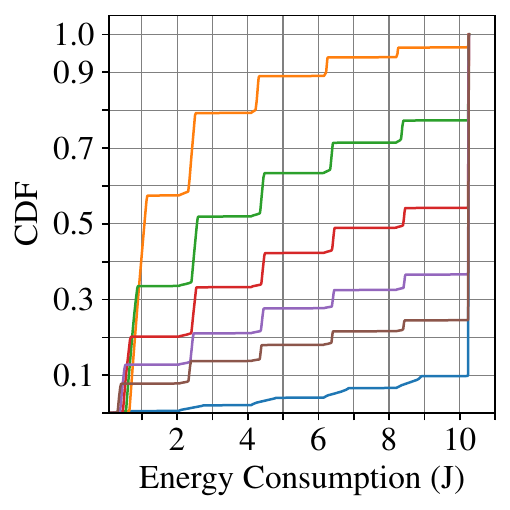}%
        \label{fig:energy_all_3gpp_4k}
    }
    \hspace{-3mm}
        \subfloat[3GPP, 5K UEs]{%
        \includegraphics[width=0.33\columnwidth]{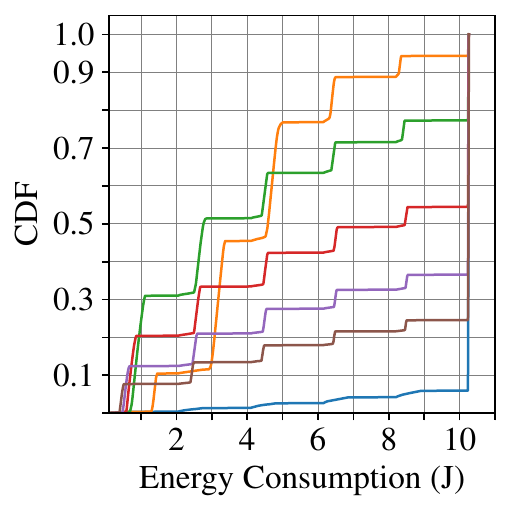}%
        \label{fig:energy_all_3gpp_5k}
    } \\[1ex]
    \caption{CDF of the energy consumption of UEs regardless of the registration result}
   \label{fig:energy_all_plots}
	\vspace*{-5mm}
\end{figure}

\begin{figure}[t]
    \centering
    \includegraphics[width=0.75\columnwidth]{Figures/legend} \hfill \\
    \vspace{-3mm}
    \subfloat[\framework{}, 3K UEs]{%
        \includegraphics[width=0.33\columnwidth]{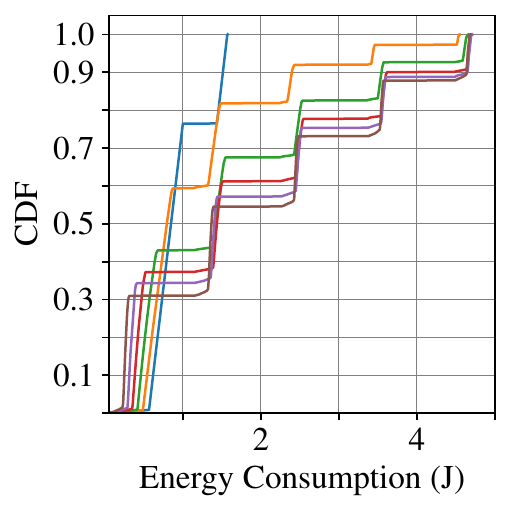}%

        \label{fig:energy_reg_astro_3k}
    }
    \hspace{-3mm}
    \subfloat[\framework{}, 4K UEs]{%
        \includegraphics[width=0.33\columnwidth]{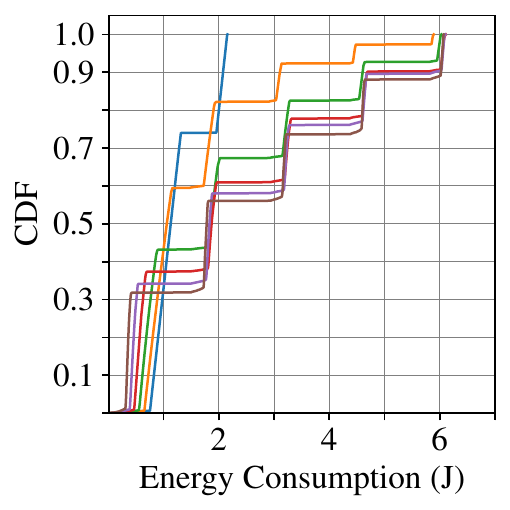}%
        \label{fig:energy_reg_astro_4k}
    }
    \hspace{-3mm}
        \subfloat[\framework{}, 5K UEs]{%
        \includegraphics[width=0.33\columnwidth]{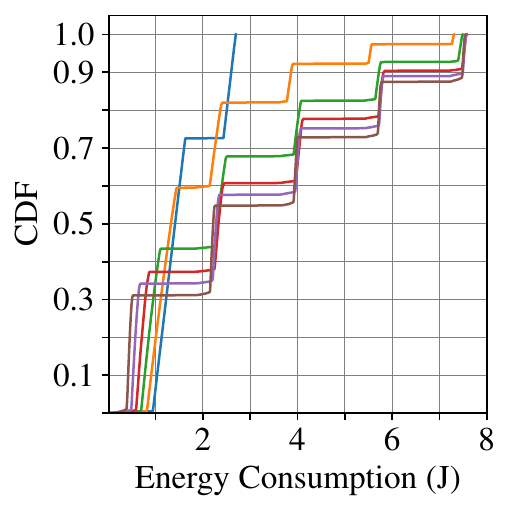}%
        \label{fig:energy_reg_astro_5k}
    } \\[1ex]
    \vspace{-2mm}
        \subfloat[3GPP, 3K UEs]{%
        \includegraphics[width=0.33\columnwidth]{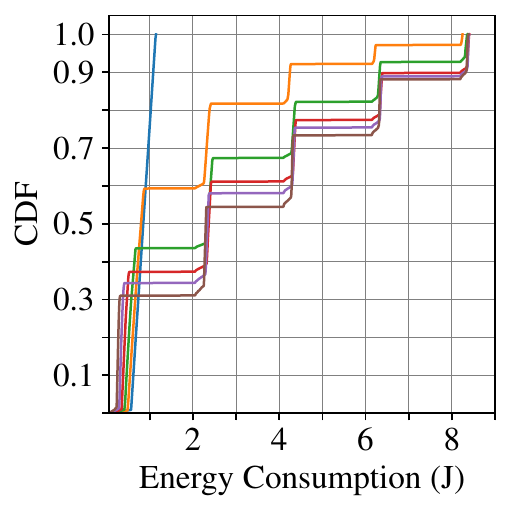}%
        \label{fig:energy_reg_3gpp_3k}
    }
    \hspace{-3mm}
    \subfloat[3GPP, 4K UEs]{%
        \includegraphics[width=0.33\columnwidth]{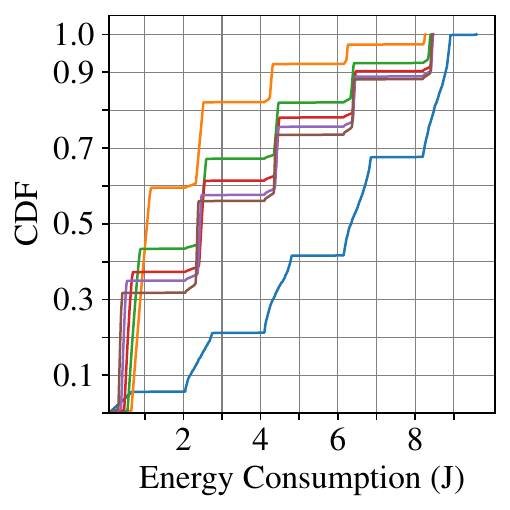}%
        \label{fig:energy_reg_3gpp_4k}
    }
    \hspace{-3mm}
        \subfloat[3GPP, 4K UEs]{%
        \includegraphics[width=0.33\columnwidth]{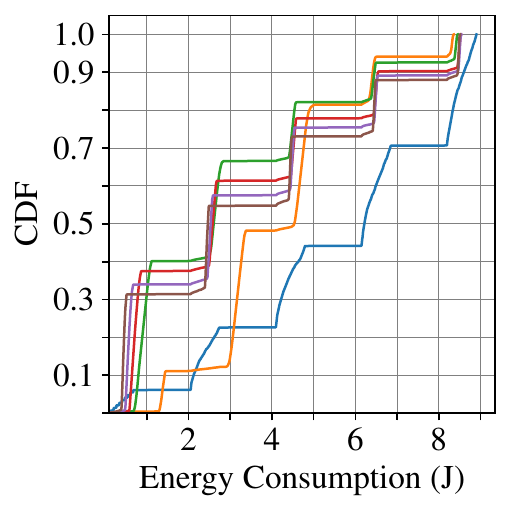}%
        \label{fig:energy_reg_3gpp_5k}
    } \\[1ex]
    \caption{CDF of the energy consumption of successfully registered UEs}
    \label{fig:energy_reg_plots}
	\vspace*{-6mm}
\end{figure}

			Here we examine the importance of timers from a power consumption perspective. \cref{fig:energy_all_plots,fig:energy_reg_plots} show the CDF of consumed energy for all (regardless of the registration outcomes) and registered UEs, respectively.
			An increase in loss probability leads to additional attempts and more time in the active state, thereby increasing energy consumption.
			\framework{} decreases energy consumption by reducing time in the active state and adapting idle time.
    \section{Conclusion}\label{sec:conclusion}
	With the unique characteristics LSNs that make current fixed NAS timers inefficient for future NF deployments, we designed \framework{}, which sizes NAS timers based on the above features. Then, we compared \framework{} against 3GPP NAS timers for MEO/GEO satellites. Although 3GPP values appear overprovisioned, our results show severe performance degradation in most scenarios. We demonstrated that \framework{} can reduce registration time under packet loss, lower the load on NFs by decreasing UE registration attempts, and save UE energy consumption. Furthermore, \framework{} may indirectly enhance user satisfaction by enabling earlier access to services, thereby improving overall network efficiency through fewer registration attempts and reduced resource wastage. Our future work includes exploring additional optimizations for NAS timers and evaluating its performance under diverse traffic patterns.

    \bibliographystyle{IEEEtran}
    \bibliography{References/references}

\end{document}